\documentclass{an}
\usepackage{times}
\usepackage{graphicx}

\renewcommand{\phi}{\varphi}

\begin{document}

\title{A Planetary System with an Escaping Mars}
\author{\'A. S\"uli\inst{1} \and R. Dvorak\inst{2}}
\institute{E\"otv\"os Lor\'and University, Department of Astronomy, P\'azm\'any P\'eter
s\'et\'any 1/A, Budapest H-1518, Hungary
\and
Institute for Astronomy, University of Vienna, T\"urkenschanzstr. 17, A-1180 Wien, Austria}
\received{}
\accepted{}
\publonline{}
\keywords{solar system: general -- instabilities -- methods: numerical}
\abstract{The chaotic behaviour of the motion of the planets in our Solar System is well
established. In this work to model a hypothetical extrasolar planetary system our Solar System
was modified in such a way that we replaced the Earth by a more massive planet and let the
other planets and all the orbital elements unchanged. 
The major result of former numerical experiments with a modified Solar System
was the appearance of a chaotic window at $\kappa_E \in (4,6)$, where the dynamical state of
the system was highly chaotic and even the body with the smallest mass escaped in some cases.
On the contrary for very large values of the mass of the Earth, even greater than that of
Jupiter regular dynamical behaviour was observed. In this paper the investigations are extended
to the complete Solar System and showed, that this chaotic window does still exist.
Tests in different 'Solar Systems' clarified that including only Jupiter and Saturn with their actual
masses together with a more 'massive' Earth ($4 < \kappa_E < 6$) perturbs the orbit of Mars so that
it can even be ejected from the system. Using the results of the Laplace-Lagrange secular theory we
found secular resonances acting between the motions of the nodes of Mars, Jupiter and Saturn.
These secular resonances give rise to strong chaos, which is the cause of the appearance
of the instability window.}
\maketitle{}

\section{Introduction}

In a previous work we studied in detail how our planets would evolve when their masses
would be different from those which they have nowadays. The primary goal was there to
study the stability
of the systems with regard to their masses without changing the actual distribution of
the orbital elements of the planets. Such systems may serve as models for exoplanetary
systems with a comparable distribution of the semimajor axes, relatively small eccentricities
and two dominating masses like Jupiter and Saturn. Additionally it can be shown with this
kind of research how terrestrial like planets would dynamically evolve in such systems.
In a first experiment (\cite{Dvorak2002} = paper I) the masses of the three terrestrial planets
Venus, Earth and Mars were uniformly enlarged and took as a dynamical model the
truncated Solar System from the planets Venus to Saturn (={\bf Ve2Sa}). It turned out that
the system remained in a dynamically stable state for enlargement factors up to more
than 200. We then started to enlarge the masses of the inner planets separately (\cite{Dvorak2005} = paper II and \cite{Suli2005} = paper III) again in the model {\bf Ve2Sa}
and found especially interesting results for the case when we
enlarged the mass of the Earth by multiplying it with the mass factor $\kappa_E$.
It turned out that the new system under consideration is stable (with quasiperiodic
motions of the planets) up to quite a large mass of the Earth. In the experiments the systems
were stable for 20 million years when the mass of the Earth was up to about 1.6-fold mass of
Jupiter. But we found a surprising exception: around $\kappa_E=5$ in all our computations
the modified Solar System is in a highly chaotic state with Mars suffering from large
eccentricities and even from escapes. In continuation of this work the interesting dynamical
behaviour is now studied in the complete Solar System including also Mercury and also the outer
two ice giants Uranus and Neptune.

\section{Methods of investigation}

As has been shown in many different studies the use of long term integration of the motions in
the planetary system gives reliable results up to at least several hundred millions of years
for a qualitative study of the orbits (e.g. \cite{Ito2002}). This means that we have
a good knowledge of the semimajor axis, the eccentricities and the inclinations of the orbits
of the planets involved. On a long term scale the motions of the planets are chaotic which was
shown by different authors (e.g. \cite{Laskar1988,Laskar1996}; \cite{Murray1999};
\cite{Lecar2001}). Additionally in a work by \cite{Laskar1994} he found that in a very far future (some
$10^9$ years) by several slight modifications of the initial conditions in the Solar System -- where the
semimajor axes were kept constant -- Mercury could get eccentricities close to 1. Already in the abstract
he claims that ''The chaotic diffusion of Mercury is so large that its eccentricity can potentially
reach values very close to 1, and ejection of this planet out of the Solar system resulting from close
encounter with Venus is possible in less than 3.5 Gyr.'' Our modification is somewhat more drasticall
but it is not our purpose to simulate our Solar System as it is but to investigate it as a special
model for extrasolar planetary systems.

For our new investigation we used a program already used and tested in many other applications
of orbit dynamics, namely the Lie-integration (e.g. \cite{Dvorak2003}; \cite{Asghari2004}).
The method is based on the integration of differential equations with Lie-series and uses the
property of recurrence formulae for the Lie-terms. This method has an automatic step size control
which makes its results reliable also for eccentric orbits, whilst no additional computations are
necessary to accomplish (in contrary to symplectic methods). The details of the method are described
in the appendix.

During the long term integrations we checked the evolution of the action like elements, which
-- in case of a chaotic orbit -- show quite irregular behaviour: small to moderate jumps in
the eccentricities and inclinations and also in the semimajor axis can be found. Finally in
many cases we found a 'quasiescape'\footnote{with orbital eccentricities $e > 0.91$} of the
planet with a small mass, i. e. Mars. What we were interested in is to investigate more
in detail the dynamical behaviour of a 'Solar-like' planetary system (with all terrestrial
planets and the ice giants) when we increase the mass of the Earth with a factor $\kappa_E$
between 4 and 6.

\section{The dynamical models}

First we have undertaken computations in different dynamical models inside this
'chaotic window': in the truncated models {\bf Ve2Ju},
{\bf Ve2Ma} and {\bf Ea2Ma} the motion of Mars did not show any signs of chaos
when we enlarged the mass of the Earth via $\kappa_E$. These carefull examinations
showed that the inclusion of Mercury did not significantly modified the dynamical
evolution of the inner planets, although it's mass is still comparable to the those of
the other three terrestrial planets. It therefore seems clear that the couple of
Jupiter-Saturn in the model {\bf Ve2Sa} is -- together with a more massive Earth --
responsible for the escapes of Mars.

The effect of Uranus and Neptune on the dynamics of the planets and on the size and
location of the chaotic window (if it is still exist) is of high concern. In order to
study the dynamics of these systems several numerical integrations were performed in the
chaotic window. To present the main features of the results the specific value of
$\kappa_E=4.7$ was selected (chosen just as one out of several others in this window).
For this mass factor the evolution of the action like elements will be shown and discussed
in details and also a comparison with the former results is given.

\subsection{The 'truncated' planetary system {\bf Ve2Sa}}

\begin{figure}
\centerline{
\includegraphics[width=0.7\linewidth,angle=270]{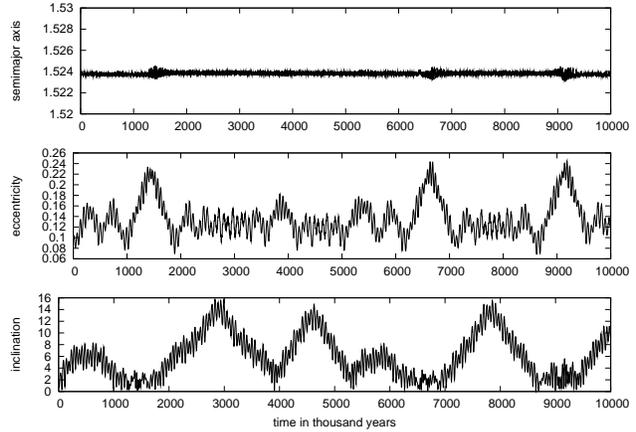}}
\caption{Orbital evolution of Mars in the dynamical model {\bf Ve2Sa} with
$\kappa_E=4.7$; the larger variations in the semimajor axis (upper plot) coincides
with larger eccentricities (middle graph) and relative small inclinations (lower graph).}
\label{fig:1}
\end{figure}

\begin{figure}
\centerline{
\includegraphics[width=0.7\linewidth,angle=270]{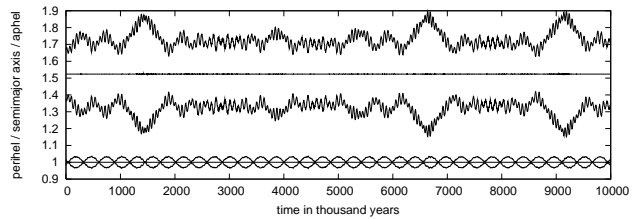}}
\caption{The time evolution of the semimajor axes the perihelia and aphelia distances
of Mars and Earth in the dynamical model {\bf Ve2Sa} with $\kappa_E=4.7$ for
10 million years.}
\label{fig:2}
\end{figure}

In Fig. \ref{fig:1} (upper panel) we can see a kind of irregular variations in the
semimajor axis of Mars; they go together with large values of the eccentricities
(middle panel). During these phases the inclination (lower panel) is always
relatively small. On the contrary when the inclination is large the variations
in the eccentricity and the value itself is relatively small. This is a consequence of
the fact that Delaunay element $H=\sqrt{a(1-e^2)}\cos i$ changes slowly with time.
In Fig. \ref{fig:2} the semimajor axis, the perihelion and aphelion distances of
Mars together with the respective orbital elements of the Earth are plotted. As one
can see the two orbits are still far from intersection, nevertheless whenever Mars
comes close to the Earth these kind of 'punches' act as larger perturbations on its
semimajor axis (see upper panel of Fig. \ref{fig:1}).

\begin{figure}
\centerline{
\includegraphics[width=0.7\linewidth,angle=270]{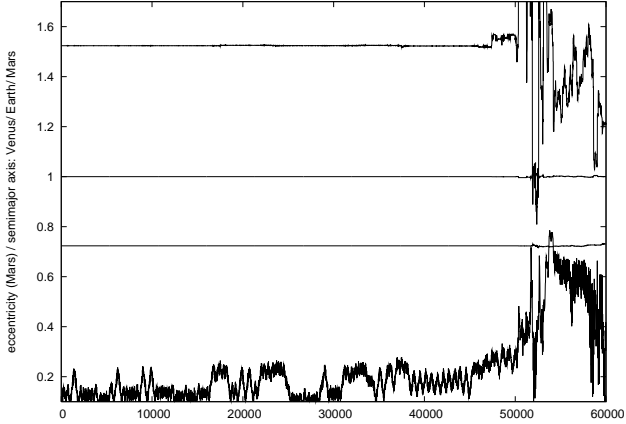}}
\caption{Orbital Evolution of the semimajor axes for the three terrestrial
planets (upper lines) and the eccentricity of Mars for 60 million years in
the model {\bf Ve2Sa} ($\kappa_E=4.7$).}
\label{fig:3}
\end{figure}

In Fig. \ref{fig:3} we see the results of a continuation of the computations shown
in Fig. \ref{fig:2} up to the moment when the eccentricities of the orbit of Mars reaches
values of $\approx$ 0.8 after 54 Myrs. From Fig. \ref{fig:2} and \ref{fig:3} it is clear that the
semimajor axis of the Earth is constantly 1.0 AU and its aphelion distance reaches periodically
1.05 AU, therefore when the eccentricity of Mars is higher than 0.31 its orbit may cross that
of the Earth. Throughout the integration this limit is approached for several short time interval.
Although the distance between the two orbits ($\approx 0.1$ AU) during these
periods are still an order of magnitude bigger than the Hill-sphere of Earth
($\approx 0.01$ AU) the Earth can strongly perturb the motion of Mars.

After 45 million years the eccentricity of Mars begins to grow secularly, reaches values
bigger than 0.31. This is followed by a cascade mechanism caused by the mutual orbital
crossings: the eccentricity of Mars suffers from very big jumps with amplitude as high as
0.75: the orbit of Mars crosses the orbits of the Earth and Venus too and it is
only a matter of time that the highly chaotic orbit leads to a subsequent escape of Mars.

\subsection{The complete system {\bf Me2Ne}}

\begin{figure}
\centerline{
\includegraphics[width=0.7\linewidth,angle=270]{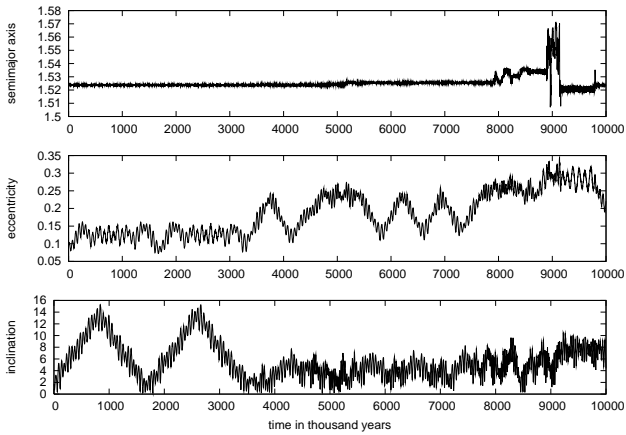}}
\caption{Evolution of the the action like variables of Mars for 10 million
years in the model {\bf Me2Ne}; semimajor axis (upper graph), eccentricity
(middle graph) and inclination (lower graph) ($\kappa_E=4.7$).}
\label{fig:4}
\end{figure}

\begin{figure}
\centerline{
\includegraphics[width=0.7\linewidth,angle=270]{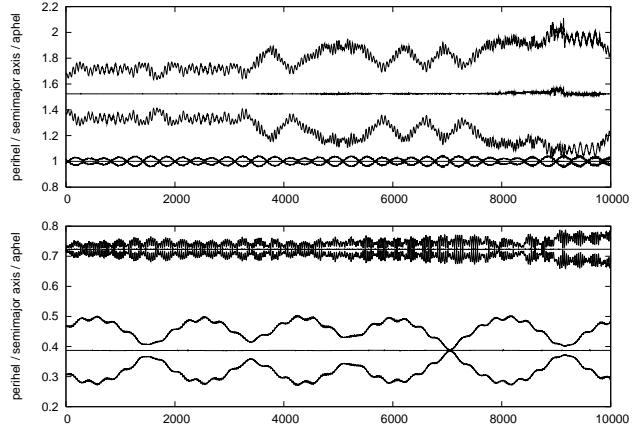}}
\caption{Orbital evolution of the semimajor axis and the perihelia and aphelia distances
for the Earth and Mars (upper panel) for 10 million years in the model {\bf Me2Ne};
the same elements are plotted for Mercury and Venus (lower panel) ($\kappa_E=4.7$).}
\label{fig:5}
\end{figure}

In Fig. \ref{fig:4} (top graph) one can observe a small increase (jump) in the semimajor
axis after almost 4 million years (because of the difference in the $y$-scale compared
to Fig. \ref{fig:1} it is not well visible but it is present) together with an increase
in the eccentricity. From the same moment on the inclination (bottom graph) stays in a
mode of only small irregular changes up to the end of integration of 10 million years.
Large irregular variations superimposed on a high mean value of $e \approx 0.26$ after
7.8 million years lead to strong variations in the semimajor axis ($1.51 < a < 1.57$)
of Mars as it can be seen in Fig. \ref{fig:4} (upper graph). In Fig. \ref{fig:5} 
(upper graph) this dynamical behaviour is well explained
when we see that the aphelion distance of the Earth and the perihelion distance of Mars
almost equals after nine millions of years. This means that Mars may enter into the
Hill-sphere of the Earth. In the same Fig. \ref{fig:5} (lower graph) one can see that also
Venus is suffering from an increase in the eccentricity. It is well visible that Mercury
does not play any important role in the dynamical evolution during the first ten million
years (lower graph in Fig. \ref{fig:5}). In both models discussed the motion of the outer
planets did not show any visible different behaviour compared to the actual Solar System.

\section{Determination of the secular frequencies}

For a possible explanation of the strong chaotic behaviour of the system for
special values of $\kappa_E$ we have applied the first order secular theory
of Laplace-Lagrange. It can be used when the eccentricities and inclinations
can be regarded as small quantities, the orbits are not crossing and no
mean motion commensurabilities are present.
It is also important that the masses involved are small with respect to the
primary body, which is for sure the case even with a 6-fold masses in the case of the
Earth (still much smaller than the gas giants). With the orbital elements:
\begin{eqnarray}
\left( \begin{array}{c}
h \\
k \\
\end{array} \right) =
e \cdot \begin{array}{c}
\sin \varpi \\
\cos \varpi \\
\end{array}
,\qquad
\left( \begin{array}{c}
p \\
q \\
\end{array} \right) =
i\cdot \begin{array}{c}
\sin \Omega \\
\cos \Omega \\
\end{array}
\label{eq:1}
\end{eqnarray}
the Laplace-Lagrange solution reads:
\begin{eqnarray}
\left( \begin{array}{c}
h_s \\
k_s \\
\end{array} \right) &=&
\sum_{j=1}^n M_s^{(j)} \begin{array}{c}
\sin \\
\cos \\
\end{array}
\left( g_jt+\beta_j\right), \label{eq:2} \\
\left( \begin{array}{c}
p_s \\
q_s \\
\end{array} \right) &=&
\sum_{j=1}^n L_s^{(j)} \begin{array}{c}
\sin \\
\cos \\
\end{array}
\left( f_jt+\gamma_j\right), \label{eq:3}
\end{eqnarray}
where $N$ is the number of bodies ($N=5$ for $\bf Ve2Sa$ and $N=8$ for $\bf Me2Ne$),
$M_s^{(j)}, L_s^{(j)}$ are the amplitudes, $g_j, f_j$ are the secular frequencies,
and $\beta_j, \gamma_j$ are the phases.
\begin{table*}
\centering
\begin{minipage}{190mm}
\caption{The $f_j$ secular frequencies, and the $L_s^{(j)}$ amplitudes of the model for $\kappa_E=5.2$
determined by the Laplace-Lagrange theory. $f_j$ in arcsec/yr.}
{\small
\begin{tabular}{|r| r r r r r r r r|}
\hline
$f_j$ & -48.769679 &-26.194843 &-26.041015 &-9.7343131 &-7.0784088 &-2.8810179 &-0.67270058 & 0.0 \\
\hline
$L_s^{(j)}$ & $j=1$ & $j=2$ & $j=3$ & $j=4$ & $j=5$ & $j=6$ & $j=7$ & $j=8$ \\
Mercury & 2.879$\cdot 10 ^{-3}$ & 1.993$\cdot 10 ^{-3}$ & 2.444$\cdot 10 ^{-3}$ & 1.616$\cdot 10 ^{-1}$ & 6.545$\cdot 10 ^{-2}$ & 1.640$\cdot 10 ^{-3}$ & 1.600$\cdot 10 ^{-4}$ & 2.796$\cdot 10 ^{-2}$ \\
Venus &-5.241$\cdot 10 ^{-2}$ &-7.126$\cdot 10 ^{-3}$ &-9.535$\cdot 10 ^{-3}$ &-2.747$\cdot 10 ^{-3}$ & 2.190$\cdot 10 ^{-2}$ & 1.259$\cdot 10 ^{-3}$ & 1.520$\cdot 10 ^{-4}$ & 2.796$\cdot 10 ^{-2}$ \\
Earth & 7.427$\cdot 10 ^{-3}$ &-3.862$\cdot 10 ^{-3}$ &-5.503$\cdot 10 ^{-3}$ &-2.975$\cdot 10 ^{-3}$ & 2.026$\cdot 10 ^{-2}$ & 1.225$\cdot 10 ^{-3}$ & 1.510$\cdot 10 ^{-4}$ & 2.796$\cdot 10 ^{-2}$ \\
Mars &-2.277$\cdot 10 ^{-3}$ & 3.867$\cdot 10 ^{-1}$ & 3.723$\cdot 10 ^{-1}$ &-1.876$\cdot 10 ^{-3}$ & 1.141$\cdot 10 ^{-2}$ & 1.051$\cdot 10 ^{-3}$ & 1.465$\cdot 10 ^{-4}$ & 2.796$\cdot 10 ^{-2}$ \\
Jupiter &-4.370$\cdot 10 ^{-6}$ &-1.875$\cdot 10 ^{-3}$ & 4.059$\cdot 10 ^{-3}$ & 9.975$\cdot 10 ^{-6}$ &-1.132$\cdot 10 ^{-4}$ & 7.469$\cdot 10 ^{-4}$ & 1.376$\cdot 10 ^{-4}$ & 2.796$\cdot 10 ^{-2}$ \\
Saturn & 2.056$\cdot 10 ^{-6}$ & 4.589$\cdot 10 ^{-3}$ &-1.017$\cdot 10 ^{-2}$ & 1.823$\cdot 10 ^{-5}$ &-1.652$\cdot 10 ^{-4}$ & 6.113$\cdot 10 ^{-4}$ & 1.327$\cdot 10 ^{-4}$ & 2.796$\cdot 10 ^{-2}$ \\
Uranus &-6.524$\cdot 10 ^{-9}$ &-1.946$\cdot 10 ^{-4}$ & 4.374$\cdot 10 ^{-4}$ &-4.657$\cdot 10 ^{-6}$ & 7.325$\cdot 10 ^{-5}$ &-1.390$\cdot 10 ^{-2}$ &-1.315$\cdot 10 ^{-4}$ & 2.796$\cdot 10 ^{-2}$ \\
Neptune & 7.922$\cdot 10 ^{-10}$&-2.170$\cdot 10 ^{-5}$ & 4.889$\cdot 10 ^{-5}$ &-4.275$\cdot 10 ^{-7}$ & 4.827$\cdot 10 ^{-6}$ & 1.642$\cdot 10 ^{-3}$ &-1.388$\cdot 10 ^{-3}$ & 2.796$\cdot 10 ^{-2}$ \\
\hline
\end{tabular}
}
\label{tab:1}
\end{minipage}
\end{table*}

To determine the first order solution of the dynamical model as a function
of the mass factor $\kappa_E$, we have computed the $g_j(\kappa_E)$, $f_j(\kappa_E)$,
$M_s^{(j)}(\kappa_E)$, $L_s^{(j)}(\kappa_E)$ and $\beta_j(\kappa_E)$, $\gamma_j(\kappa_E)$
functions for $\kappa_E \in [4,6]$ with the aid of the {\sc MAPLE} algebra manipulation package. 
The comparison of the results from our determination up to the first order with 
Bretagnon 1974, 1982 and Kne\v zevic 1986 showed satisfactory agreement for the
model $\bf Me2Ne$ (with $\kappa_E=1$).

The orbital elements of the $s$th planet are described by Eq. (\ref{eq:2}) and Eq. (\ref{eq:3}),
which are the sum of harmonic oscillations. Using these formulae it can be calculated
that the planets' eccentricities and inclinations are varying between given limits
with quasiperiodic oscillations. Due to the positive $g_j$ secular angular velocities
the apsidal lines of the planets are rotating in the same direction as the planets, 
whereas the nodes accordingly to the negative $f_j$ secular angular velocities
(see the first line of Table \ref{tab:1}) are are rotating in the opposite direction. Upon these
mean rotations quasiperiodic variations are superimposed. Both the apsidal and
nodal motions can be approximated by average angular velocities, which are to a first
approximation equal with the frequencies of those harmonious terms which are multiplied
by the largest amplitudes:
\begin{eqnarray}
e_s \cdot \begin{array}{c}
\sin \varpi_s\\
\cos \varpi_s\\
\end{array}
\approx
M_s^{(J)} \begin{array}{c}
\sin \\
\cos \\
\end{array}
\left( g_Jt+\beta_J\right), \label{eq:4} \\
i_s \cdot \begin{array}{c}
\sin \Omega_s\\
\cos \Omega_s\\
\end{array}
\approx
L_s^{(K)} \begin{array}{c}
\sin \\
\cos \\
\end{array}
\left( f_Kt+\gamma_K\right), \label{eq:5}
\end{eqnarray}
where $M_s^{(J)} = \mathbf{max}_j |M_s^{(j)}|,\,L_s^{(K)} = \mathbf{max}_j |L_s^{(j)}|$
and the average angular velocities of the $s$th planet are given by $g_J$ and $f_K$.
In this manner the secular frequencies can be associated with each planet. We note that
this assignment is not unambiguous.

\begin{figure}
\centerline{
\includegraphics[width=0.9\linewidth,angle=0]{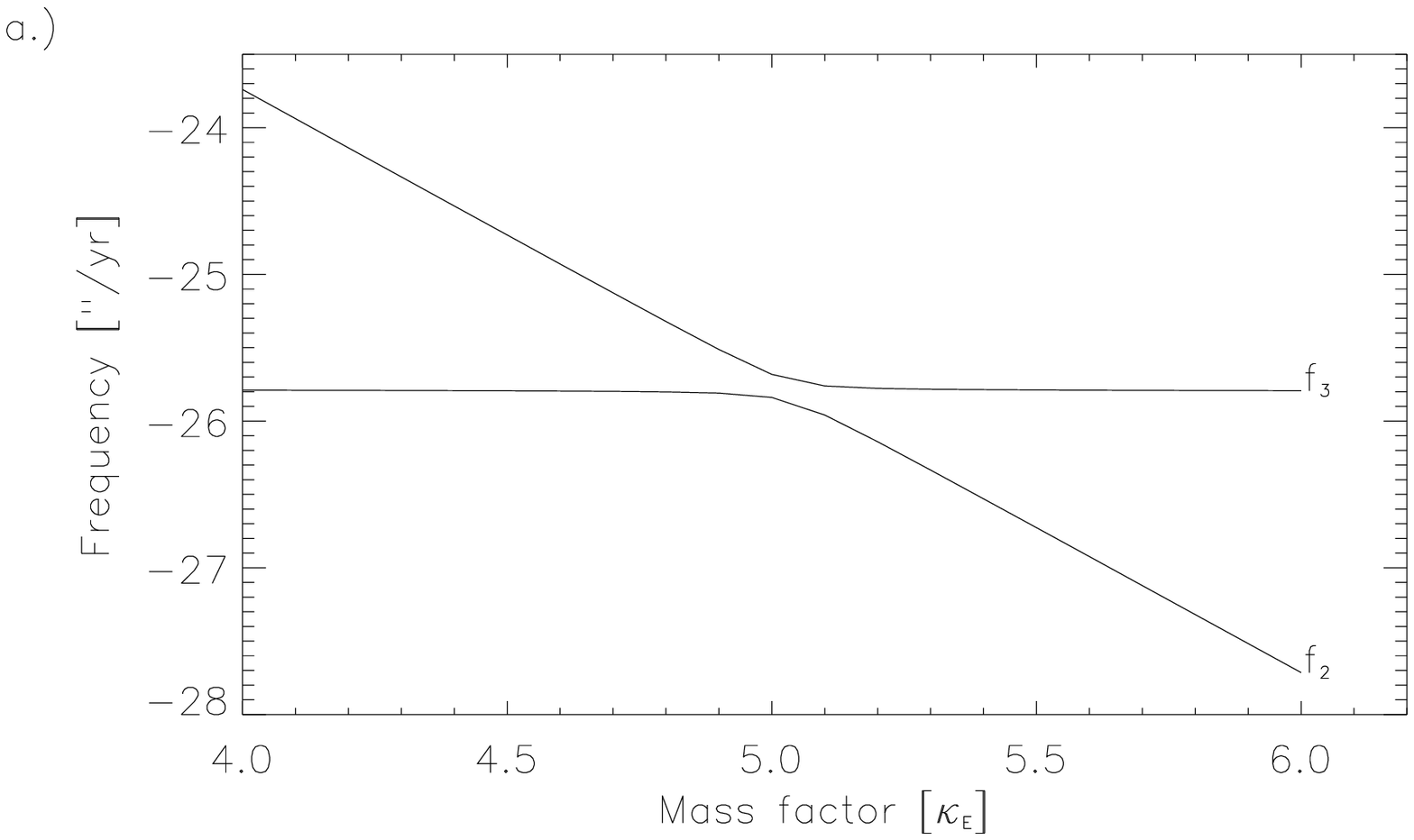}}
\centerline{
\includegraphics[width=0.9\linewidth,angle=0]{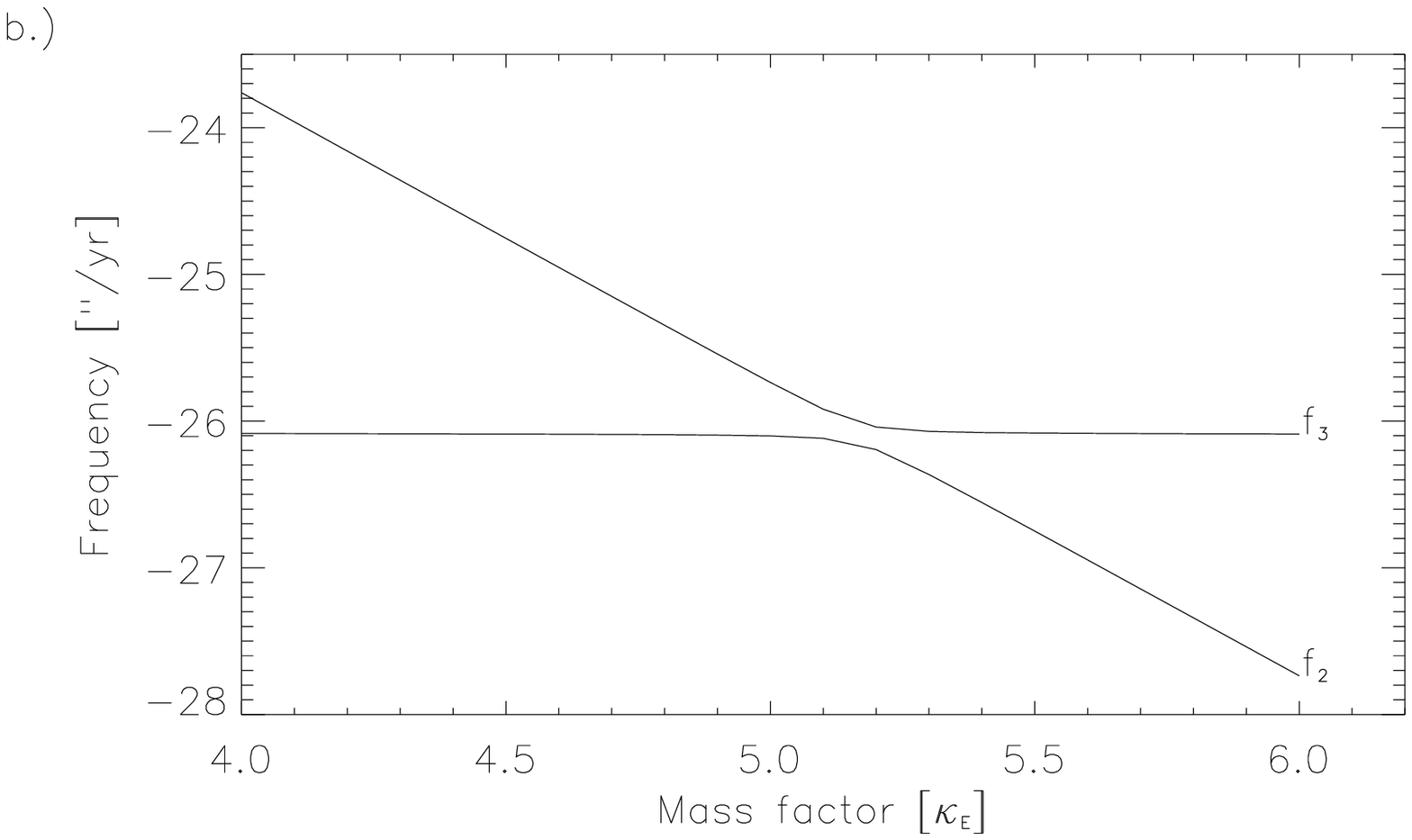}}
\caption{Secular frequencies $f_j$ in the model {\bf Ve2Sa} (a)
and {\bf Me2Ne} (b) with respect to the mass factor $\kappa$.}
\label{fig:6}
\end{figure}

In Fig. \ref{fig:6} we compare the frequencies $f_2$ and $f_3$ assigned to the planets Earth
and Mars, respectively, in the models {\bf Ve2Sa} and {\bf Me2Ne}, which show only a small
shift along the $\kappa_E$ axis: the minimum distance is 0.1567 arcsec/year 
for $\kappa_E \approx 5.00$
in the model {\bf Ve2Sa} and 0.1538 arcsec/year for $\kappa_E \approx 5.20$ in the model {\bf Me2Ne}

A study of Table \ref{tab:1} shows that the largest amplitude are, in the
solution for Mars, $L_4^{(2)}$ and $L_4^{(3)}$, for Jupiter $L_5^{(2)}$
and $L_5^{(3)}$ and for Saturn $L_6^{(2)}, L_6^{(3)}$. Accordingly the $f_2$
and $f_3$ frequencies, can be associated with Mars, Jupiter and Saturn.
The orbital plane of Mars therefore on the average rotates together with those
of Jupiter and Saturn, giving rise to chaotic behaviour. The equality of two
apsidal or nodal rates is referred to in Solar System as a secular resonance.
In this case we have three secular resonances:
$\dot \Omega_M \approx \dot\Omega_J$, $\dot\Omega_M \approx
\dot\Omega_S$ and $\dot\Omega_J \approx \dot\Omega_S$.
We suspect that these secular resonances are the main
source of the observed chaos, and produce the chaotic window.

\section{Discussion}

\begin{figure}
\centerline{
\includegraphics[width=0.38\linewidth,angle=270]{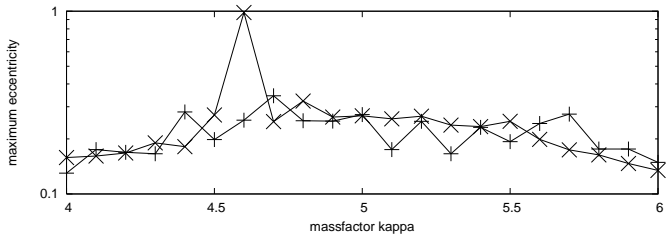}}
\caption{Comparison of the maximum value of the eccentricity of Mars in the models
{\bf Ve2Sa} ($\times$) and {\bf Me2Ne} (+) for 10 million years with respect to the
mass factor $\kappa_E$.}
\label{fig:7}
\end{figure}

\begin{figure}
\centerline{
\includegraphics[width=0.7\linewidth,angle=270]{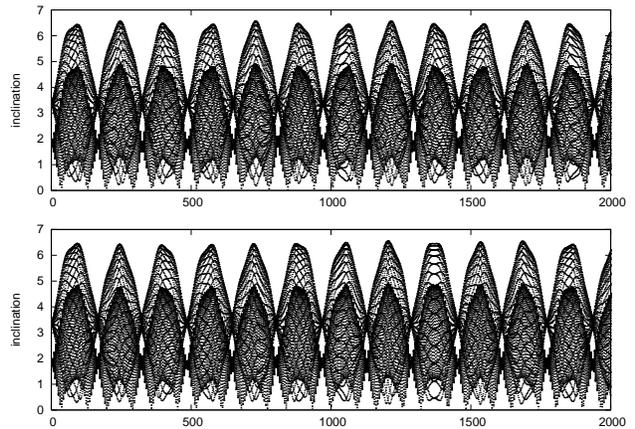}}
\caption{Comparison of the inclinations of the planets Venus and Mars in the
models {\bf Ve2Sa} (upper graph) and {\bf Me2Ne} (lower graph) for the Earth with
a mass of Saturn for 2 million years.}
\label{fig:8}
\end{figure}

In modelling extrasolar planetary systems we took our Solar System as starting
point for several models, where we increased the masses of the planets
involved until to the moment of instability. In former papers it turned out that
only a significantly larger mass of the Earth (the other planets' masses were left
unchanged) could lead to a decay of these modified systems. In the article we focus
on a surprising ``chaotic window'' in the dynamical evolution of our planetary system
when we increase the mass of the Earth by the massfactor $4<\kappa_E<6$. Former results
of paper III have unveiled this interesting dynamical behaviour for a truncated Solar
system model ($\bf Ve2Sa$). We used a heuristic way to find out the reason for this
unexpected strong chaotic behaviour: we numerically integrated different dynamical models.
It turned out that only the couple Jupiter-Saturn together with the Earth (with a larger
mass) is the cause for the subsequent escape of Mars. In the comparison of the two models
{\bf Ve2Sa} with {\bf Me2Ne} we see that there is in principle no difference for the state of
chaoticity of the orbits: in both models Mars suffers sooner or later from close
approaches with the Earth because of the large values of the eccentricity (Fig. \ref{fig:7}).
This is true for the whole interval $4 \le \kappa_E \le 6$ which we tested with a step of
$\Delta \kappa_E = 0.1$ for both models. On the contrary, with larger values of
$\kappa_E < 540 \approx 1.8 M_{Jupiter}$ a very regular dynamical behaviour was observed for all planets.
This regular dynamical evolution can be seen in the
quasiperiodic behaviour of the inclinations of the planets (Fig. \ref{fig:8}). In the upper
panel one can see the inclinations of Venus (max($i$)=$6.^{\circ}5$) and Mars
(max($i$)$=5^{\circ}$) for
the model {\bf Ve2Sa}, in the lower panel the same quantities in the model {\bf Me2Ne}.
It is evident that qualitatively both plots agree quite well. The first quantitative
differences appear after 1 million years in the inclination of Venus. The same overall
behaviour can be observed for the eccentricities (not shown). All these orbits are stable.
But why do we have strong chaos in this window of $\kappa$ which does not appear for
larger values up to a mass factor which correspond to an Earth comparable to Jupiter?
Using the results of the Laplace-Lagrange secular theory we found secular resonances
acting between the motions of the nodes of Mars, Jupiter and Saturn. These secular resonances
give rise to strong chaos, which is the primary cause of the appearance of the chaotic window,
and eventually the escape of Mars. The properties of the dynamics of the model in the chaotic
window must be further analysed by higher order secular theory.
The final answer to this problem is highly interesting for future research on the
dynamics of extrasolar planetary systems especially when we will have evidence
via observations -- primarely from space missions like KEPLER, DARWIN and
TPF-- that some of them are also hosting terrestrial planets.

\acknowledgements
We thank the Wissenschaftlich-technisches \\ Abkommen \"Osterreich-Ungarn Project A12-2004.
Most of the numerical integrations were accomplished on the NIIDP (National Information
Infrastructure Development Program) supercomputer in Hungary. This study was supported
by the International Space Science Institute (ISSI) and benefit from the ISSI team
'Evolution of Habitable Planets'.

\appendix

\section{The Lie-integration}

Because our integration method is not one which is used by many colleagues we
shortly explain how the LIE-integration works. This method, based on an idea by
\cite{Grobner1967}, has been used by our group since 1984
(e.g. \cite{Hanslmeier1984}; \cite{Lichtenegger1984}; Delva 1984,1985;
\cite{Dvorak1993}). It turned out to be a
precise and fast tool, which can be also used when the orbits of a system
suffer from close encounters. This is done by computing for every step the
optimal step size for the desired precision of the integration.

Let $D$ denote a linear differential operator; the point $z=(z_1,z_2,\dots,z_n)$ lies
in the n-dimensional z-space; the functions $\theta_i(z)$ are holomorphic
within a certain domain $G$, e.g. they can be expanded in converging power
series. Let the function $f(z)$ be holomorphic in the same region as
$\theta_i(z)$. Then $D$ can be applied to $f(z)$:
\begin{equation}
Df=\theta_1(z)\frac{\partial f}{\partial z_1} + \theta_2(z)\frac{\partial f}{\partial z_2}
+ \dots + \theta_n(z)\frac{\partial f}{\partial z_n}
\label{def}
\end{equation}
If we proceed applying $D$ to $f$ we get
\begin{eqnarray}
D^2f&=&D(Df) \nonumber \\
\vdots&& \nonumber \\
D^nf&=&D(D^{n-1}f) \nonumber
\end{eqnarray}
The {\bf Lie-series} will be defined in the following way;
\begin{displaymath}
L(z,t)=\sum_{\nu=0}^{\infty} \frac{t^\nu}{\nu!} D^\nu f(z) = f(z)+ tDf(z) +
\frac{t^2}{2!}D^2f(z)+\dots
\end{displaymath}
Because we can write the Taylor-expansion of the exponential function
\begin{equation}
e^{tD}f = 1 + tD^1 + \frac{t^2}{2!}D^2+\frac{t^3}{3!}D^3+\dots
\end{equation}
$L(z,t)$ can be written in the symbolic form
\begin{equation}
L(z,t)=e^{tD}f(z)
\end{equation}
The convergence proof of $L(z,t)$ is given in detail in \cite{Grobner1967}.
The most useful property of Lie-series is the {\sf Vertauschungssatz}:
\begin{theorem}
Let $F(z)$ be a holomorphic function in the neighbourhood of
$(z_1,z_2,\dots,z_n)$ where the corresponding power series expansion
converges at the point $(Z_1,Z_2,\dots,Z_n)$; then we have:
\label{ver}
\end{theorem}
\begin{equation}
F(Z)= \sum_{\nu=0}^{\infty} \frac{t^\nu}{\nu!} D^\nu F(Z)
\end{equation}
or
\begin{equation}
F(e^{tD})z=e^{tD}F(z)
\end{equation}
Making use of it we can demonstrate how Lie-series solve differential equations. Let us give the
system of differential equations:
\begin{equation}
\frac{dz_i}{dt}=\theta_i(z)
\label{des}
\end{equation}
with $(z_1,z_2,\dots,z_n)$. We postulate that the solution of~(\ref{des}) can be written as
\begin{equation}
z_i=e^{tD}\xi_i
\label{res}
\end{equation}
where $\xi_i$ are the initial conditions $z_i(t=0)$ and D is the
Lie-operator as defined in~(\ref{def}). In order to prove~(\ref{res}) we
differentiate it with respect to time $t$ and make use of the {\sf Vertauschungssatz}:
\begin{equation}
\frac{dz_i}{dt}=De^{tD}\xi_i=e^{tD}D\xi_i.
\end{equation}
Because of
\begin{equation}
D\xi_i=\theta_i(\xi_i)
\end{equation}
we obtain -- again by using the {\sf Vertauschungssatz} -- the following result which
turns out to be the original differential equation~(\ref{des}):
\begin{equation}
\frac{dz_i}{dt}=e^{tD}\theta_i(\xi_i)=\theta_i(e^{tD}\xi_i)=\theta_i(z_i)
\end{equation}

\end{document}